\documentclass[12pt]{article}%
\usepackage{amsmath}
\usepackage{amsfonts}
\usepackage{amssymb}
\usepackage{graphicx}%
\begin{document}

\title{Recovering the equivalence of ensembles}
\author{Vera B. Henriques and Silvio R. Salinas\\Institute of Physics,\\University of S\~{a}o Paulo\\S\~{a}o Paulo, SP, Brazil}
\date{16 January, 2015}
\maketitle

\begin{abstract}
The equivalence of thermodynamic results in the canonical and the
microcanonical ensembles has been questioned in some calculations for spin
models with long-range interactions. We show that these claims of
inequivalence are related to an inadequate definition of the independent
(density) variables in the microcanonical ensemble. We illustrate this point
with the example of a simple spin-$1$ ideal paramagnet, and then revisit the
original calculations of Barr\'{e}, Mukamel, and Ruffo, for a mean-field
spin-$1$ Blume-Capel model. If the microcanonical ensemble is defined in terms
of adequate density variables, we show that there is no disagreement with the
calculations in the usual canonical ensemble (with fixed thermodynamic field variables).

\end{abstract}

The equivalence of ensembles, which is one of the hallmarks of modern
statistical physics \cite{ruelle1969}\cite{griffiths1972}, has been challenged
by Barr\'{e}, Mukamel and Ruffo \cite{barre2001}, in the calculations for a
simple mean-field, fully connected, spin-$1$ Hamiltonian. This work motivated
a number of remarks on similar thermodynamic discrepancies between results in
the usual canonical and microcanonical ensembles associated with spin model
systems with long-range interactions \cite{campa2009}.

We show that this claimed inequivalence of ensembles is due to an inadequate
definition of the independent variables in the microcanonical ensemble. A
proper definition of the microcanonical ensemble is illustrated by analytic
calculations for a spin-$1$ ideal paramagnet. We then revisit the original
work of Barr\'{e} and collaborators for a mean-field version of the spin-$1$
Blume-Capel model, which is known to display a temperature-anisotropy phase
diagram with second and first-order transition lines and a tricritical point.
If the microcanonical ensemble is defined in terms of the adequate density
variables, we show that there is no disagreement with the well-known solutions
in the usual canonical ensemble (with fixed temperature and thermodynamic
field variables). According to Griffiths and Wheeler \cite{griffiths1970}, we
distinguish between \textquotedblleft density\textquotedblright\ variables,
given by the ratio of thermodynamic extensive quantities, and
\textquotedblleft field\textquotedblright\ variables, which \textquotedblleft
take on identical values in two phases that are in thermodynamic equilibrium
with each other\textquotedblright.

\section{Ideal paramagnet of spin-1}

In this Section, we use a spin-$1$ ideal paramagnet to give a very simple
example of the choice of variables in the usual canonical and the
microcanonical formalisms. Consider a spin Hamiltonian given by%
\begin{equation}
\mathcal{H}=-H\sum_{j=1}^{N}S_{j}+D\sum_{j=1}^{N}S_{j}^{2},
\end{equation}
where $H$ and $D$ are \textquotedblleft field variables\textquotedblright, and
$S_{j}=-1,0,+1$, for $j=1,2,...,N$.

\subsection{Calculations in the canonical ensemble}

In the canonical ensemble (with fixed values of the field variables, $H$, $D$,
and temperature $T$), we have the partition function%
\[
Z=Z\left(  T,H,D\right)  =\sum_{\left\{  S_{j}\right\}  }\exp\left[  \beta
H\sum_{j=1}^{N}S_{j}-\beta D\sum_{j=1}^{N}S_{j}^{2}\right]  =
\]%
\begin{equation}
=\left[  1+2\exp\left(  -\beta D\right)  \cosh\left(  \beta H\right)  \right]
^{N},
\end{equation}
where $\beta=1/k_{B}T$. The associated thermodynamic potential per site is
given by%
\begin{equation}
g=g\left(  T,H,D\right)  =-\frac{1}{\beta}\ln\left[  1+2\exp\left(  -\beta
D\right)  \cosh\left(  \beta H\right)  \right]  .
\end{equation}
This is a fundamental equation \cite{callen1960} in the $T-H-D$
representation. From this thermodynamic potential, we obtain the equations of
state,%
\begin{equation}
m=m\left(  T,H,D\right)  =-\left(  \frac{\partial g}{\partial H}\right)
_{T,D}=\frac{\sinh\left(  \beta H\right)  }{\frac{1}{2}\exp\left(  \beta
D\right)  +\cosh\left(  \beta H\right)  }, \label{param}%
\end{equation}
where the thermodynamic density $m$ is the dimensionless magnetization per
site, $m=\left\langle \sum_{j=1}^{N}S_{j}\right\rangle /N$, and%
\begin{equation}
q=q\left(  T,H,D\right)  =\left(  \frac{\partial g}{\partial D}\right)
_{T,H}=\frac{\cosh\left(  \beta H\right)  }{\frac{1}{2}\exp\left(  \beta
D\right)  +\cosh\left(  \beta H\right)  }, \label{paraq}%
\end{equation}
where the density $q$ is a dimensionless \textquotedblleft
quadrupole\textquotedblright, $q=\left\langle \sum_{j=1}^{N}S_{j}%
^{2}\right\rangle /N$. The entropy per site is given by the remaining equation
of state,%
\[
s=s\left(  T,H,D\right)  =-\left(  \frac{\partial g}{\partial T}\right)
_{H,D}=k_{B}\ln\left[  1+2\exp\left(  -\beta D\right)  \cosh\left(  \beta
H\right)  \right]  +
\]%
\begin{equation}
+k_{B}\frac{\beta D\,\cosh\left(  \beta H\right)  }{\frac{1}{2}\exp\left(
\beta D\right)  +\cosh\left(  \beta H\right)  }-\frac{\beta H\,\sinh\left(
\beta H\right)  }{\frac{1}{2}\exp\left(  \beta D\right)  +\cosh\left(  \beta
H\right)  }. \label{parascanonical}%
\end{equation}
In this usual canonical representation, the thermodynamic densities $m$, $q$,
and $s$, are obtained as functions of the fields $T$, $H$, and $D$. The
meaning of $D$ as a thermodynamic field can be fully appreciated if we use the
spin-$1$ model to mimic a binary lattice gas of particles, in which case $D$
is related to a chemical potential.

It is now interesting to analyze some features of this ideal spin-$1$
paramagnet. Using equations (\ref{param}) and (\ref{paraq}), the entropy can
be written as%
\begin{equation}
s=-\frac{1}{T}g+\frac{D}{T}q-\frac{H}{T}m, \label{paraeuler}%
\end{equation}
which shows that, in this schematic model, the field terms in the effective
Hamiltonian do not contribute to the \textquotedblleft magnetic internal
energy\textquotedblright. In other words, the average value of the spin
Hamiltonian is a kind of \textquotedblleft magnetic enthalpy\textquotedblright%
, which cannot be taken as the internal energy \cite{callen1960}. Equation
(\ref{paraeuler}) is just the analogous of the Euler relation for a simple
fluid system. It is then straightforward to use this Euler relation, and the
expression for the free energy, $g=g\left(  T,H,D\right)  $, to write the
differential form%
\begin{equation}
ds=\frac{D}{T}dq-\frac{H}{T}dm,
\end{equation}
which shows that the entropy per particle $s$ can also be written as a
function of the densities $m$ and $q$. In this special case, we can perform
the appropriate calculations to write the analytic expression%
\[
s=s\left(  q,m\right)  =-k_{B}\left(  1-q\right)  \ln\left(  1-q\right)
-k_{B}\frac{1}{2}\left(  q-m\right)  \ln\left(  q-m\right)  -
\]%
\begin{equation}
-k_{B}\frac{1}{2}\left(  q+m\right)  \ln\left(  q+m\right)  +k_{B}q\ln2,
\label{paraentropy}%
\end{equation}
which is equivalent to a fundamental equation in the microcanonical ensemble
(in terms of the densities $m$ and $q$, and the number of sites $N$; note the
fixed value, $u=0$, of the internal magnetic energy in this simple
paramagnetic model). Incidentally, this expression has the same form as the
\textquotedblleft entropy to be maximized\textquotedblright, given by equation
(5) in the article of Barr\'{e} and collaborators \cite{barre2001}.

\subsection{Calculations in the microcanonical ensemble}

We now show that the same expression for the entropy per particle, $s=s\left(
q,m\right)  $, given by equation (\ref{paraentropy}), can also be obtained in
the context of a properly defined microcanonical ensemble, in terms of the
densities $m$ and $q$.

In this simple problem, the number of microscopic configurations is given by%
\begin{equation}
\Omega=\Omega\left(  N_{+},N_{0},N_{-}\right)  =\frac{N!}{N_{+}!N_{0}!N_{-}!},
\label{microstates}%
\end{equation}
where $N$ is the total number of spins, and $N_{+}$, $N_{0}$, and $N_{-}$, are
the numbers of spins up, zero, and down, respectively. Taking into account
that
\begin{equation}
N=N_{+}+N_{0}+N_{-}, \label{number}%
\end{equation}
and that,%
\begin{equation}
m=\frac{1}{N}\left(  N_{+}-N_{-}\right)  ,\qquad q=\frac{1}{N}\left(
N_{+}+N_{-}\right)  , \label{mandq}%
\end{equation}
we can write $\Omega$ as a function of $N$, $m$, and $q$. In the thermodynamic
limit, we show that
\begin{equation}
s=s\left(  q,m\right)  =\lim_{N\rightarrow\infty}\frac{1}{N}k_{B}\ln\Omega,
\label{paraentropymicro}%
\end{equation}
where $s=s\left(  q,m\right)  $ is a fundamental equation in the entropy
representation, given by the same expression, equation (\ref{paraentropy}),
which we have already obtained in the context of the canonical ensemble. From
this form of the entropy, we obtain the equations of state in the entropy
representation,%
\begin{equation}
\frac{D}{T}=\left(  \frac{\partial s}{\partial q}\right)  _{m},\qquad\frac
{H}{T}=-\left(  \frac{\partial s}{\partial m}\right)  _{q},
\end{equation}
from which we regain the usual expressions of $m$ and $q$ in the canonical
ensemble, given by equations (\ref{param}) and (\ref{paraq}).

At this point we have shown the full equivalence between the canonical and the
microcanonical formulations of this simple problem. We can also resort to an
alternative (and more general) formulation in the microcanonical ensemble,
which turns out to be useful for dealing with short-range interacting systems
as well.

In the alternative formulation, we rewrite the canonical partition function in
the form%
\[
Z=Z\left(  T,H,D\right)  =\sum_{\left\{  S_{j}\right\}  }\int dm\int
dq\,\delta\left(  m-\frac{1}{N}\sum_{j=1}^{N}S_{j}\right)  \delta\left(
q-\frac{1}{N}\sum_{j=1}^{N}S_{j}^{2}\right)  \times
\]%
\begin{equation}
\times\exp\left(  \beta HNm-\beta DNq\right)  ,
\end{equation}
which can also be written as%
\begin{equation}
Z=\int dm\int dq\,\Omega\left(  q,m,N\right)  \,\exp\left(  \beta HNm-\beta
DNq\right)  ,
\end{equation}
where $\Omega\left(  q,m,N\right)  $ is the number of microstates with fixed
values of $q$, $m$, and $N$. We now introduce convenient integral
representations of the delta functions. In the thermodynamic limit, we have%
\begin{equation}
\Omega\sim\exp\left[  Nf\left(  q,m;k_{1},k_{2}\right)  \right]  ,
\end{equation}
with%
\begin{equation}
f=k_{1}m+k_{2}q+\ln\left[  1+2\exp\left(  -k_{2}\right)  \cosh\left(
k_{1}\right)  \right]  , \label{paraf}%
\end{equation}
where the parameters $k_{1}$ and $k_{2}$ come from the saddle-point equations,%
\begin{equation}
\frac{\partial f}{\partial k_{1}}=\frac{\partial f}{\partial k_{2}}=0.
\end{equation}
The microcanonical entropy is given by $s=s\left(  q,m\right)  =k_{B}f$, with
$k_{1}$ and $k_{2}$ obtained from the extremization of the function $f\left(
q,m;k_{1},k_{2}\right)  $. It is not difficult to show that $k_{1}=-\beta H$,
and $k_{2}=\beta D$, so that equation (\ref{paraf}) leads to the same
expression of equation (\ref{parascanonical}). Again, we recover the known
results in the canonical ensemble.

\section{The spin-1 Blume-Capel model}

We now consider the special spin system analyzed by Barr\'{e} and
collaborators \cite{barre2001}, given by the fully connected spin-$1$
Hamiltonian%
\begin{equation}
\mathcal{H}=-\frac{J}{2N}\left(  \sum_{j=1}^{N}S_{j}\right)  ^{2}+D\sum
_{j=1}^{N}S_{j}^{2}, \label{bch}%
\end{equation}
where $J$ and $D$ are positive parameters, and $S_{j}=+1,0,-1$ for all sites.
In the canonical ensemble (fixed values of the field variables $T$ and $D$),
we have%
\begin{equation}
Z=Z\left(  T,D,N\right)  =\sum_{\left\{  S_{j}\right\}  }\exp\left[
\frac{\beta J}{2N}\left(  \sum_{j=1}^{N}S_{j}\right)  ^{2}-\beta D\sum
_{j=1}^{N}S_{j}^{2}\right]  . \label{bcz}%
\end{equation}
Using a Gaussian identity, we write%
\begin{equation}
Z=\left(  \frac{\beta JN}{2\pi}\right)  ^{1/2}\int_{-\infty}^{+\infty}%
dy\,\exp\left[  -\beta N\,g\left(  T,D;y\right)  \right]  , \label{bcz2}%
\end{equation}
where%
\begin{equation}
g\left(  T,D;y\right)  =\frac{J}{2}y^{2}-\frac{1}{\beta}\ln\left[
1+2\exp\left(  -\beta D\right)  \cosh\left(  \beta Jy\right)  \right]  .
\label{bcg}%
\end{equation}
In the thermodynamic limit, the free energy $g=g\left(  T,D\right)  $ comes
from the minimization of $g\left(  T,D;y\right)  $ with respect to $y$. We
then have%
\begin{equation}
g\left(  T,D\right)  =g\left(  T,D;\widetilde{y}\right)  , \label{bcg2}%
\end{equation}
with%
\begin{equation}
\widetilde{y}=\frac{\sinh\left(  \beta J\widetilde{y}\right)  }{\frac{1}%
{2}\exp\left(  \beta D\right)  +\cosh\left(  \beta J\widetilde{y}\right)  },
\label{bcy}%
\end{equation}
and the proviso about the existence of multiple solutions for $\widetilde{y}$
as a function of the thermodynamic fields $T$ and $D$.

We now expand $g\left(  T,D;y\right)  $, given by equation (\ref{bcg}), as a
power series of $y$, with field-dependent coefficients. From this Landau-like
expansion, we obtain the well-known results for the critical line in the $D-T$
space,%
\begin{equation}
D=k_{B}T\ln\left[  \frac{2J}{k_{B}T}-2\right]  , \label{critical}%
\end{equation}
with $1/3<k_{B}T/J<1$, and a tricritical point at $k_{B}T/J=1/3$ (and
$D/J=\left(  \ln4\right)  /3$). For $k_{B}T/J<1/3$, there is a line of
first-order transitions, with the coexistence of a paramagnetic and an ordered
phase at well-defined values of the thermodynamic fields $T$ and $D$. As it is
usual in the mean-field models, the location of this first-order border comes
from a proper application of a Maxwell construction.

Some simple manipulations lead to the usual equations of state, for the
entropy and the quadrupole moment, in the $T-D$ representation,%
\begin{equation}
q=\left(  \frac{\partial g}{\partial D}\right)  _{T}=\frac{\cosh\left(  \beta
J\widetilde{y}\right)  }{\frac{1}{2}\exp\left(  \beta D\right)  +\cosh\left(
\beta J\widetilde{y}\right)  }, \label{bccanq}%
\end{equation}
and%
\begin{equation}
s=-\left(  \frac{\partial g}{\partial T}\right)  _{D}=k_{B}\ln\left[
1+2\exp\left(  -\beta D\right)  \cosh\left(  \beta J\widetilde{y}\right)
\right]  +\frac{D}{T}q-\frac{J}{T}\widetilde{y}^{2}, \label{bccanentropy}%
\end{equation}
where $\widetilde{y}$ comes form the solutions of equation (\ref{bcy}). It
easy to see that $\widetilde{y}$ corresponds to the magnetization $m$ per
site, $\left\langle \sum S_{j}\right\rangle /N$, but we have to be careful in
the region of multiple solutions (and first-order transitions).

\subsection{Calculations for the Blume-Capel model in the microcanonical
ensemble}

As in the work of Barr\'{e}, Mukamel, and Ruffo \cite{barre2001}, we initially
write the number of microstates,%
\begin{equation}
\Omega=\Omega\left(  N_{+},N_{0},N_{-}\right)  =\frac{N!}{N_{+}!N_{0}!N_{-}!},
\end{equation}
where $N$ $=N_{+}+N_{0}+N_{-}$ is the fixed number of spins. We now fix the
quadrupole density,%
\begin{equation}
q=\frac{N_{+}+N_{-}}{N},
\end{equation}
and the internal magnetic energy per site,%
\begin{equation}
u=-\frac{J}{2N^{2}}\left(  N_{+}-N_{-}\right)  ^{2}=-\frac{J}{2}\left(
\frac{N_{+}-N_{-}}{N}\right)  ^{2}, \label{bcu}%
\end{equation}
and write $\Omega$ as a function of $N$, $u$, and $q$. In the thermodynamic
limit, the entropy is given by
\begin{equation}
s=s\left(  u,q\right)  =\lim_{N\rightarrow\infty}\frac{1}{N}k_{B}\ln\Omega.
\label{bcmicros}%
\end{equation}

In this particular mean-field model, the internal energy $u$ is associated
with the magnetization $m$,%
\begin{equation}
u=-\frac{J}{2}\left(  \frac{N_{+}-N_{-}}{N}\right)  ^{2}=-\frac{J}{2}m^{2},
\label{bcu2}%
\end{equation}
so that fixed values of $u$ also correspond to fixed values of $m^{2}$.
Therefore, the entropy $s=s\left(  u,q\right)  $ is given by%
\[
s=s\left(  u,q\right)  =-k_{B}\left(  1-q\right)  \ln\left(  1-q\right)
-k_{B}\frac{1}{2}\left[  q-m\right]  \ln\left[  q-m\right]  -
\]%
\begin{equation}
-k_{B}\frac{1}{2}\left[  q+m\right]  \ln\left[  q+m\right]  +k_{B}q\ln2,
\label{bcentropy}%
\end{equation}
which is identical to equation (\ref{paraentropy}), for the ideal spin-$1$
paramagnet, but we now have to take into account that $m$ is a function of $u
$, $m=m\left(  u\right)  $, according to the definition of the internal energy
of this mean-field model, given by equation (\ref{bcu2}). Note that $u<0$.
Also, note the symmetry with respect to $\pm m$.

Given the fundamental equation for this system, $s=s\left(  u,q\right)  $, we
write the equations of state in the entropy representation,%
\begin{equation}
\frac{1}{T}=\left(  \frac{\partial s}{\partial u}\right)  _{q},\qquad\frac
{D}{T}=\left(  \frac{\partial s}{\partial q}\right)  _{u}.
\end{equation}
It is straightforward to show that these equations of state lead to the same
expressions for the internal energy $u$ (related to the magnetization $m$) and
the quadrupole moment $q$, as a function of $T$ and $D$, which we have already
obtained in the context of the usual canonical ensemble. Numerical
calculations for the first-order transition border just confirm these
findings. In the region of multiple solutions of the mean-field equations,
Maxwell%
\'{}%
s construction should be properly used to choose the minima of the canonical
free energy in terms of $D$ and $T$ (which is equivalent to recovering the
convexity of the entropy in the microcanonical ensemble). In contrast to the
work of Barr\'{e}, Mukamel, and Ruffo \cite{barre2001},these calculations
indicate that there is full equivalence of ensembles.

\subsection{Alternative formulation of the microcanonical ensemble}

The equivalence of ensembles can also be shown if we resort to an alternative,
and more general formulation of the microcanonical ensemble. In this
formulation, we introduce two delta functions to fix the energy $u$ and the
quadrupole density $q$, and rewrite the canonical partition function of the
Blume-Capel model,%
\[
Z=\sum_{\left\{  S_{j}\right\}  }\int du\int dq\,\delta\left(  \left[
u+\frac{J}{2N}\left(  \sum_{j=1}^{N}S_{j}\right)  ^{2}\right]  \right)
\delta\left(  q-\frac{1}{N}\sum_{j=1}^{N}S_{j}^{2}\right)  \times
\]%
\begin{equation}
\times\exp\left[  \beta Nu-\beta DNq\right]  =\int du\int dq\,\Omega\left(
u,q,N\right)  \exp\left[  -\beta Nu-\beta DNq\right]  ,
\end{equation}
so that $\Omega\left(  u,q,N\right)  $ is the number of microstates with fixed
values of $u$, $q$, and $N$.

Using an integral representation for the delta functions, it easy to write%
\[
\Omega\left(  u,q,N\right)  =%
{\displaystyle\int\limits_{-i\infty}^{+i\infty}}
\frac{dk_{1}}{2\pi i}%
{\displaystyle\int\limits_{-i\infty}^{+i\infty}}
\frac{dk_{2}}{2\pi i}\exp\left[  Nuk_{1}+Nqk_{2}\right]  \times
\]%
\begin{equation}
\times\left(  \frac{k_{1}JN}{2\pi}\right)  ^{1/2}%
{\displaystyle\int\limits_{-\infty}^{+\infty}}
dy\exp{\LARGE \{}-\frac{1}{2}k_{1}JNy^{2}+N\ln\left[  1+2\exp\left(
-k_{2}\right)  \cosh\left(  k_{1}Jy\right)  \right]  {\LARGE \}.}%
\end{equation}
In the thermodynamic limit, we have%
\begin{equation}
\Omega\left(  u,q,N\right)  \sim\exp\left[  Nf\left(  u,q;y,k_{1}%
,k_{2}\right)  \right]  ,
\end{equation}
with%
\begin{equation}
f\left(  u,q;y,k_{1},k_{2}\right)  =k_{1}u+k_{2}q-\frac{1}{2}k_{1}Jy^{2}%
+\ln\left[  1+2\exp\left(  -k_{2}\right)  \cosh\left(  k_{1}Jy\right)
\right]  ,
\end{equation}
where the parameters $k_{1}$, $k_{2}$ and $y$ come from the saddle-point
equations,%
\begin{equation}
\frac{\partial f}{\partial k_{1}}=\frac{\partial f}{\partial k_{2}}%
=\frac{\partial f}{\partial y}=0.
\end{equation}
The microcanonical entropy is given by $s=s\left(  u,q\right)  =k_{B}f$, from
which we write the equations of state in the entropy representation,
$1/T=\left(  \partial s/\partial u\right)  _{q}$ and $D/T=\left(  \partial
s/\partial q\right)  _{u}$. Using these equations, we have%
\begin{equation}
u=-\frac{J}{2}\widetilde{y}^{2},\qquad\widetilde{y}=\frac{\sinh\left(  \beta
J\widetilde{y}\right)  }{\frac{1}{2}\exp\left(  \beta D\right)  +\cosh\left(
\beta J\widetilde{y}\right)  },
\end{equation}
and%
\begin{equation}
q=\frac{\cosh\left(  \beta J\widetilde{y}\right)  }{\frac{1}{2}\exp\left(
\beta D\right)  +\cosh\left(  \beta J\widetilde{y}\right)  },
\end{equation}
which are the known expressions in the canonical ensemble.

\section{Conclusions}

In conclusion, we have shown that claims of inequivalence of ensembles in a
class of mean-field spin models can be attributed to an inadequate definition
of the independent (density) variables in the microcanonical formalism. As an
illustration of an adequate formulation, we performed some analytic
calculations for a spin-$1$ ideal paramagnet. We then revisited the
calculations of Barr\'{e}, Mukamel, and Ruffo \cite{barre2001}. If the
microcanonical ensemble is defined in terms of the adequate density variables,
we have shown that there is no disagreement with the well-known solutions in
the usual canonical ensemble (with fixed thermodynamic field variables).


\begin{thebibliography}{9}                                                                                                %


\bibitem {ruelle1969}D. Ruelle, Statistical Mechanics: Rigorous Results, W. A.
Benjamin, Inc., New York, 1969.

\bibitem {griffiths1972}R. B. Griffiths, Rigorous Results and Theorems, in
Phase Transitions and Critical Phenomena, ed. C. Domb and M. S. Green, vol. 1,
Academic Press, New York, 1972.

\bibitem {barre2001}J. Barr\'{e}, D. Mukamel, and S. Ruffo, Phys. Rev. Lett
\textbf{87}, 030601, 2001.

\bibitem {campa2009}A. Campa. T. Dauxois, and S. Ruffo, Phys. Rep.
\textbf{480}, 57, 2009.

\bibitem {griffiths1970}R. B. Griffiths and J. C. Wheeler, Phys. Rev. A
\textbf{2}, 1047, 1970.

\bibitem {callen1960}H. B. Callen, Thermodynamics, John Wiley and Sons, New
York, 1960.
\end{thebibliography}
\end{document}